\documentclass[twocolumn,showpacs,preprintnumbers,amsmath,amssymb]{revtex4-1}
\usepackage{graphicx}% Include figure files
\usepackage{dcolumn}% Align table columns on decimal point
\usepackage{bm}% bold math
\usepackage{bbm}
\def\R{{\mathbb{R}}}

\newcommand{\be}{\begin{equation}}
\newcommand{\beq}{\begin{equation}}
\newcommand{\en}{\end{equation}}
\newcommand{\eeq}{\end{equation}}
\newcommand{\bea}{\begin{eqnarray}}
\newcommand{\ena}{\end{eqnarray}}
\newcommand{\hbo}{\hbox to 1 true cm {\hfill } }

\newcommand{\e}{\mathrm{e}}
\newcommand{\lb}{\langle \kern-.17em \langle} 
\newcommand{\rb}{\rangle \kern-.17em \rangle }
\begin{document}

%\preprint{ }

\title{The density of states approach to dense quantum systems}
% Force line breaks with \\

\author{Kurt Langfeld$^{a}$}
%\email{kurt.langfeld@plymouth.ac.uk}
\author{Biagio Lucini$^b$}

\affiliation{%
\bigskip
$^a$School of Computing \& Mathematics,
Plymouth, PL4 8AA, UK  }

\affiliation{%
$^b$College of Science, Swansea University, Swansea,  SA2 8PP,  UK
}%

\date{ March 22, 2014
%\today
}% It is always \today, today,

\begin{abstract}
We develop a first-principle generalised density of state method for studying
numerically quantum field theories with a complex action.
As a proof of concept, we show that with our approach we can solve
numerically the strong sign problem of the $Z_3$ spin model at finite
density. Our results are confirmed by standard simulations of the
theory dual to the considered model, which is free from a sign problem.
Our method opens new perspectives on {\em ab initio} simulations of
cold dense quantum systems, and in particular of Yang-Mills theories
with matter at finite densities, for which  Monte Carlo based
importance sampling are unable to produce sufficiently accurate results.
\end{abstract}

%Use showkeys class option if keyword
\maketitle

Monte Carlo simulations~\cite{Creutz:1980zw} of the theory regularised
on a lattice~\cite{Wilson:1974sk} are key for obtaining first principle
results in Quantum Chromo Dynamics
(QCD)~\cite{Gregory:2009hq} and in other strongly interacting systems,
like for instance correlated electrons in solid state
physics~\cite{RevModPhys.73.33}. 
Monte Carlo simulations rely on importance sampling, which exposes the
configurations that dominate the partition function. Importance
sampling requires a real positive Gibbs factor. Because of this restriction, many crucial problems
in  physics that could in principle have been addressed by numerical
simulations have remained unexplored. In particular,
quantum systems with matter at finite densities, among which is
cold and dense baryon matter, are described by a complex action.
The corresponding Monte Carlo simulations are hampered by the notorious
sign problem, which limits severely the applicability of this method.

%\medskip
In recent years, there has been noticeable progress in numerical
studies of complex action
systems, both with Monte Carlo methods and techniques that do not rely
on importance sampling. Among the most promising methods are the complexification
of the fields in a Langevin based approach~\cite{Aarts:2011zn,Aarts:2012ft}, 
worm or flux algorithms~\cite{Prokof'ev:2001zz,Alet:2004rh}  to simulate the dual theory when the
corresponding duality transformation is known and exposes a real
action~\cite{Mercado:2012ue,Langfeld:2013kno} and the use of techniques
that explicit exploit the cancellations of classes of  fields~\cite{Chandrasekharan:2010iy}. 

%\medskip 
Among alternative approaches to conventional Monte Carlo sampling, an
efficient strategy relies on the numerical computation of the density
of states~\cite{Wang:2001ab}. Once this quantity has been determined, the partition function and
derived expectation values of observables can be computed semi-analytically,
integrating numerically the density of states with the appropriate Boltzmann weight.
An effective technique for computing the density of states for
systems with a continuous spectrum has been discussed
in~\cite{Langfeld:2012ah,Langfeld:2013xbf}. 
A natural question is whether this method, referred to as the LLR
algorithm, not relying on action-based importance samplings, could be
effective at simulating systems with a sign problem. In this letter,
we show that  a density of state approach in the LLR formulation
appropriately generalised to complex action systems can provide a
viable solution to the sign problem. As a test case to demonstrate the
method, we study the $Z_3$ spin model for finite 
chemical potentials $\mu $. This system, which has been studied also
with complex Langevin techniques~\cite{Aarts:2011zn}, provides an ideal
benchmark test for our approach, 
since it possesses a ``strong''sign problem but can be simulated with flux
type algorithms after dualisation~\cite{Mercado:2011ua}. We will show
that  our method (which  does not rely on a the existence of a dual
theory with real action, but is formulated using the original degrees of
freedom), can achieve reliable results for a wide range of
$\mu $. 

%\medskip 
%
\begin{figure*}
\includegraphics[height=7cm]{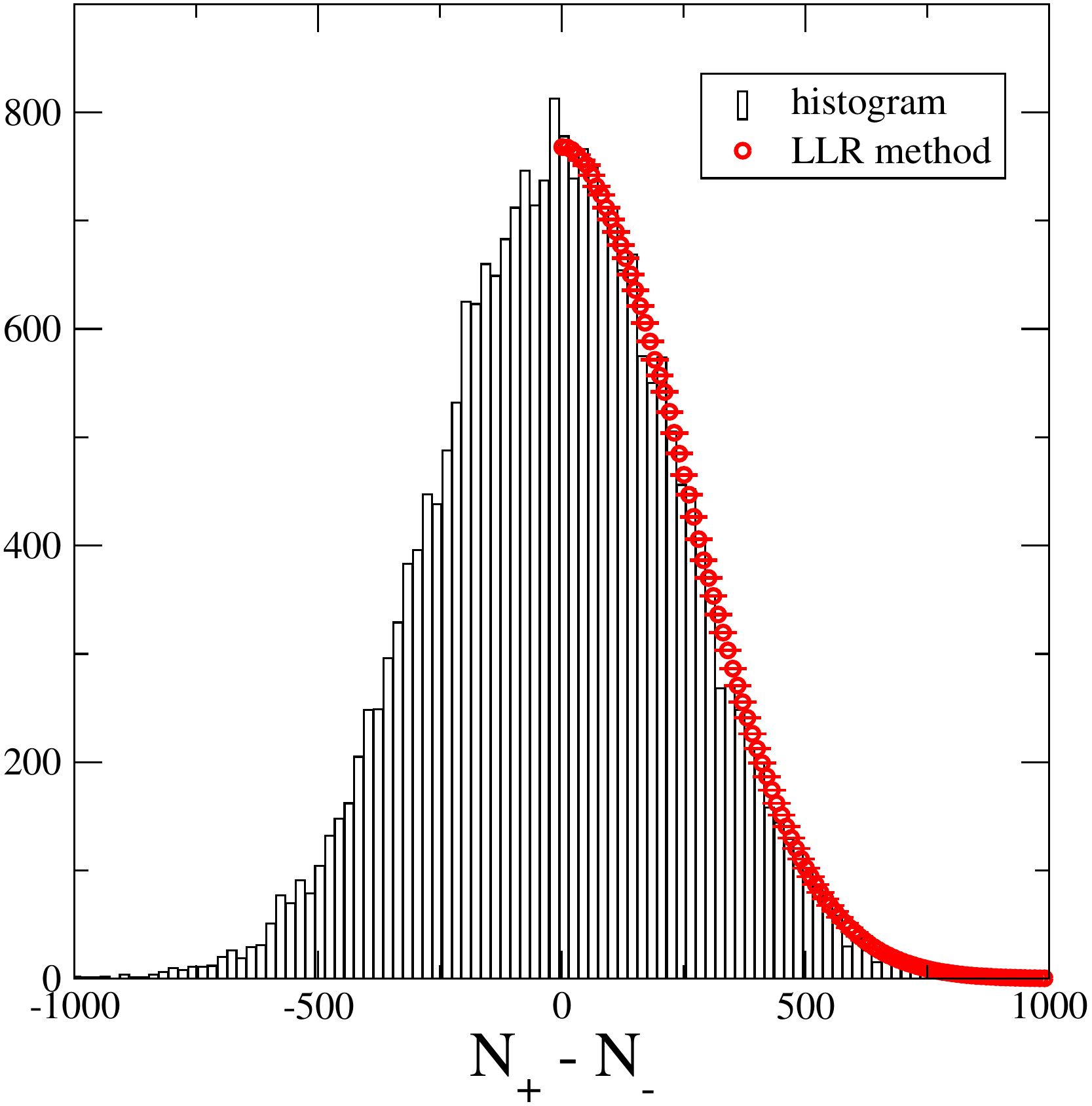} \hspace{1cm}
\includegraphics[height=7cm]{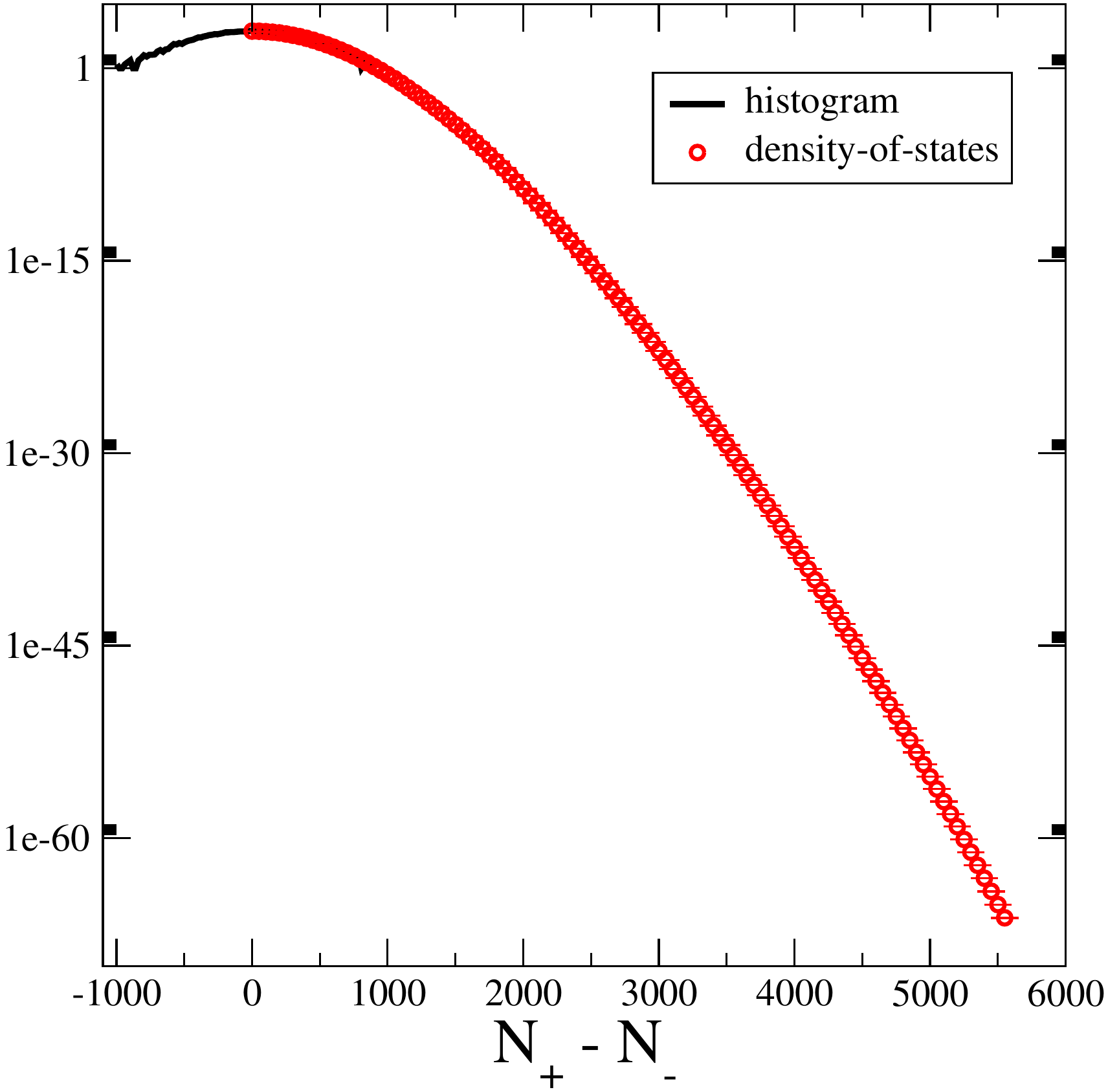}
\caption{\label{fig:1} The probability distribution $\rho $ for $n=\Delta N =
  N_+ - N_-$ 
from a direct simulation using a histogram (black steps) and from our LLR 
method (left). Same probability distribution on a logarithmic scale
(right), for a wider range of $n$. $24^3$ lattice, $\tau =0.17$, and $\kappa
= 0.05$. 
}
\end{figure*}
Before discussing in detail the considered model and our
solution technique, we shall outline how the relevant quantities (i.e. the generalised density of
states and observables sensitive to strong cancellations) are identified in a more general
setup. We consider a quantum field theory (QFT) with a complex
action. In general terms, the partition function of such a system is
given in terms of a functional integral over the degrees of 
freedom $\phi (x)$: 
\be 
Z(\mu ) \; = \; \int {\cal D} \phi \; \exp \Bigr\{ i S_I[\phi ](\mu) \Bigl\}
\;  \e ^{S_R[\Phi ](\mu )} \; , 
\label{eq:1}
\en 
with $S_R , \, S_I \in \R $ and where $\mu $ is the chemical
potential. In finite density QFTs, the imaginary part vanishes with vanishing
$\mu $, i.e., $S_I(\mu ) \to 0 $ for $\mu \to 0$. 
The simplest way to deal with the sign problem is to adopt a ``quenched''
approximation and to ignore the phase factor. This is undoubtedly a good
approximation at small $\mu$, but most likely it will fail when density effects start to play a
significant role. To quantify the importance of the phase factor, we introduce 
\be 
Z_\mathrm{mod} (\mu ) \; = \; \int {\cal D} \phi \; 
 \e ^{S_R[\Phi ](\mu )} \; . 
\label{eq:2}
\en 
We point out that observables of the modified theory can be easily estimated
by standard importance sampling methods. 
If we succeed to calculate the phase factor expectation value 
\be 
Q(\mu) \; = \; \frac{ Z (\mu )}{ Z_\mathrm{mod} (\mu ) } 
\; = \; \Bigl\langle \exp \Bigl\{ i S_I[\phi ](\mu) \Bigr\} \Bigr\rangle
_\mathrm{mod} ,  
\label{eq:3}
\en 
observables such as the density $\sigma $ would be accessible as well: 
\be 
\sigma (\mu) \; = \; \frac{d \, \ln Z }{d\mu } \; = \; 
\frac{d \, \ln Q(\mu) }{d\mu } \; + \; \frac{d \, \ln Z_\mathrm{mod} }{d\mu } 
\; . 
\label{eq:4}
\en 
Our strategy to calculate $Q(\mu )$ is the based upon the density of state
method originally proposed by Wang and Landau~\cite{Wang:2001ab} 
in its LLR version~\cite{Langfeld:2012ah}. At the heart of our approach is the
generalised density of states $\rho (s)$: 
\be 
\rho (s) \; = \; N \; \int {\cal D} \phi \; \delta \Bigl( s \, - \, 
S_I[\phi](\mu) \, \Bigr) \;  \e ^{S_R[\Phi ](\mu )} \; .  
\label{eq:5}
\en 
Later, we will choose the normalisation $N$ such that $ \rho (0)=1$. 
The phase factor can be then obtained by calculating two integrals: 
\be 
Q(\mu ) \; = \; \frac{ \int ds \; \rho (s) \; \exp \{is \} }{ \; 
\int ds \; \rho (s) } \; . 
\label{eq:6}
\en 
Note that the normalisation $N$ drops out. The challenge is that 
for sizeable and phenomenological interesting values of $\mu $ the phase
factor can be very small ($Q \approx 10^{-16}$ in the example below) and 
exponentially depends on the system volume. The smallness of $Q$ arises from 
cancellations in the numerator of (\ref{eq:6}). On the other hand, $\rho (s)$ 
is at times of order one and only known numerically. Thus, any algorithm which
addresses $\rho (s)$ must feature an exponential error suppression in order to
muster enough precision to obtain a sensible result upon the integration in
(\ref{eq:6}). As we detail below, the LLR algorithm is just delivering that. 

%\medskip 
For a showcase of our approach, we are going study the $Z_3$ spin model at
finite chemical potential $\mu $: 
The degrees of freedom  $\phi (x) \in Z_3$ are associated with the sites of
the $N^3$ 3-dimensional lattice. The partition function and the action
of the system are given by 
\bea
Z(\mu) &=& \sum _{\{\phi\}} \; \exp \Bigl\{ S[\phi] + S_h[\phi] \Bigr\} \; , 
\label{eq:10} \\
S[\phi] &=& \tau \sum _{x,\nu } \phi_x \, \phi^\ast _{x+\nu} \; , \; \; 
S_h[\phi] = 
\sum_x \, \Bigl( \eta \phi_x + \bar{\eta } \phi^\ast _x \Bigr) \; , 
\label{eq:11} 
\ena 
with $ \eta = \kappa \, \mathrm{e}^{\mu } $ and $ \bar{\eta }  = \kappa \,
\mathrm{e}^{- \mu } $. The model can be derived from QCD in the heavy quark
and strong coupling limit~\cite{Karsch:1985cb,DeGrand:1983fk}. Thereby, 
$\kappa $ is related to the quark hopping constant, and $\mu $ is the
chemical potential. For  
$\mu = {\cal O}(1)$, this theory possesses a strong sign problem
in the above formulation. However, the reformulation 
of this model with dual variables is real (even at finite $\mu $) and can be
effectively simulated using flux type algorithms~\cite{Mercado:2011ua}. This
makes this theory an ideal benchmark test for the LLR approach. 

%\medskip 
Before showing our numerical findings, we briefly detail the
calculation of the phase factor using the flux algorithm developed by
Gattringer at al.~\cite{Mercado:2011ua}. The partition function can be
expressed in terms of dual variables $\phi _D$: 
\be 
Z(\mu ) \; = \; \sum _{\{ \phi_D \} } M(\mu, \phi_D) \; P(\phi_D) \; . 
\label{eq:32} 
\en
$Z(\mu)$ can be computed in terms of $Z(0)$. However, a simplistic
approach to this calculation will be affected by a so-called overlap
problem, whereby a partition function is sampled using configurations
derived from a statistical sampling in principle related, but in
practice with different dominant contributions.
To resolve the overlap problem, we adopt a variant of the snake
algorithm~\cite{deForcrand:2000fi}. We firstly observe that
\bea 
\frac{ Z(\mu + \Delta \mu ) }{ Z(\mu ) } &=&
\frac{1}{ Z(\mu) } \sum _{\{ \phi_D \} } \frac{ M(\mu + \Delta \mu, \phi_D) }{
    M(\mu,  \phi_D) } \, \times 
\label{eq:33} \\ 
&&  M(\mu, \phi_D) \; P(\phi_D) 
= \left\langle \frac{ M(\mu + \Delta \mu, \phi_D) }{
    M(\mu,  \phi_D) } \right\rangle _\mu \; .
\nonumber 
\ena
The latter expectation value can be efficiently evaluated with the flux
algorithm. The partition function is then obtained from: 
\be 
Z( k \, \Delta \mu ) \; = \; Z(0) \, \prod _{i=1}^k \frac{ Z(i \Delta \mu ) }{ 
 Z((i-1) \Delta \mu ) } \; ,
\label{eq:34}
\en 
with each factor $Z(i \Delta \mu ) /Z((i-1) \Delta \mu )$ evaluated
with the snake algorithm. The same approach is repeated for the
``quenched'' partition function  
$Z_\mathrm{mod} $, and the phase factor is finally obtained from 
\be 
Q(k \, \Delta \mu ) \; = \; Z( k \, \Delta \mu ) / Z_\mathrm{mod} ( k \,
\Delta \mu  ) \; . 
\label{eq:35}
\en 

%\medskip 
To proceed with our method, we introduce the centre elements  
\be 
\phi \; \in \; \{ 1, z, z^\dagger \} , \hbo z := \frac{1}{2} +
\frac{\sqrt{3}}{2} \,   i \; . 
\label{eq:12} 
\en
The linear term of the action can then be written 
\bea  
S_h[\phi] &=& \kappa \sum_{x}   \Bigl[ \mathrm{e}^{\mu } \, \phi(x) 
\; + \; \mathrm{e}^{-\mu } \, \phi^\dagger (x) \Bigr] 
\nonumber \\ 
&=& \kappa \left[ \left( 2 \, N_0 - N_z - N_{z^\ast } \right) \,
  \hbox{cosh}(\mu) \right.
\nonumber \\ 
&+& \left. i \sqrt{3} \, (N_z - N_{z^\ast}) \,
  \hbox{sinh}(\mu) \right] \; ,  
\label{eq:13}
\ena 
where $N_0$, $N_z$ and $N_{z^\ast}$ are the numbers of time-like links 
equaling a particular centre element, i.e. 
\bea 
N_0 &=& \sum_x \delta \Bigl(\phi(x),1 \Bigr) , \; \; \; 
N_z = \sum_x \delta \Bigl(\phi(x),z \Bigr) , 
\nonumber \\ 
N_{z^\ast} &=& \sum_x \delta \Bigl(\phi(x),z^\ast \Bigr) . 
\label{eq:14}
\ena 
The probability distribution for the variable $\Delta N := N_z - N_{z^\ast}$
is symmetric around zero. Thus, the partition function is real and given by 
\bea 
Z(\mu) &=& \sum _{\{\phi\} } \exp \Bigl\{ S[\phi]  \; + \; 
 \kappa  \left( 3N_0 - V \right) \, \hbox{cosh}(\mu)
 \Bigr\} 
\nonumber \\ 
&& \cos \Bigl( \sqrt{3} \, \kappa \, \Delta N \,  \hbox{sinh}(\mu)
 \Bigr) \; ,   
\label{eq:15} 
\ena
where we have used the constraint 
\be 
N_0 \; + \; N_z \; + \;  N_{z^\ast } \; = \; N^3 \; := \; V \; . 
\label{eq:16} 
\en
For a fixed lattice volume $V$, we now define the density of states
$\rho $ by  
\bea 
\rho (n) \; := \; \sum _{\{\phi\}} &&
\delta \Bigl( n, \Delta N[\phi] \Bigr) \; \; 
\exp \Bigl\{ S[\phi]  
\nonumber \\ 
&+&  \kappa  \Bigl( 3N_0 [\phi] - V \Bigr) \, \hbox{cosh}(\mu)
 \Bigr\} \; . 
\label{eq:17} 
\ena 
With this definition, the partition function can be written as a simple sum: 
\be 
Z(\mu) \; = \; \sum_{ n} \; \rho ( n) \; \cos \Bigl( \sqrt{3} \, \kappa
\,   \hbox{sinh}(\mu) \; n \Bigr) \; . 
\label{eq:18} 
\en
Using a standard Monte Carlo simulation and casting the observed 
values $\Delta N$ into an histogram would only provide enough precision to
calculate the partition function for very small values of $\mu $. 
Nevertheless, this histogram provides first insights into $\rho (n)$ and 
later will serve as an important crosscheck for any more elaborate method. 
Our results for a $24^3$ lattice using $\tau = 0.17 $ and $\kappa = 0.05$ are
shown in figure~\ref{fig:1}. 

%\medskip
%
\begin{figure}
\includegraphics[height=7cm]{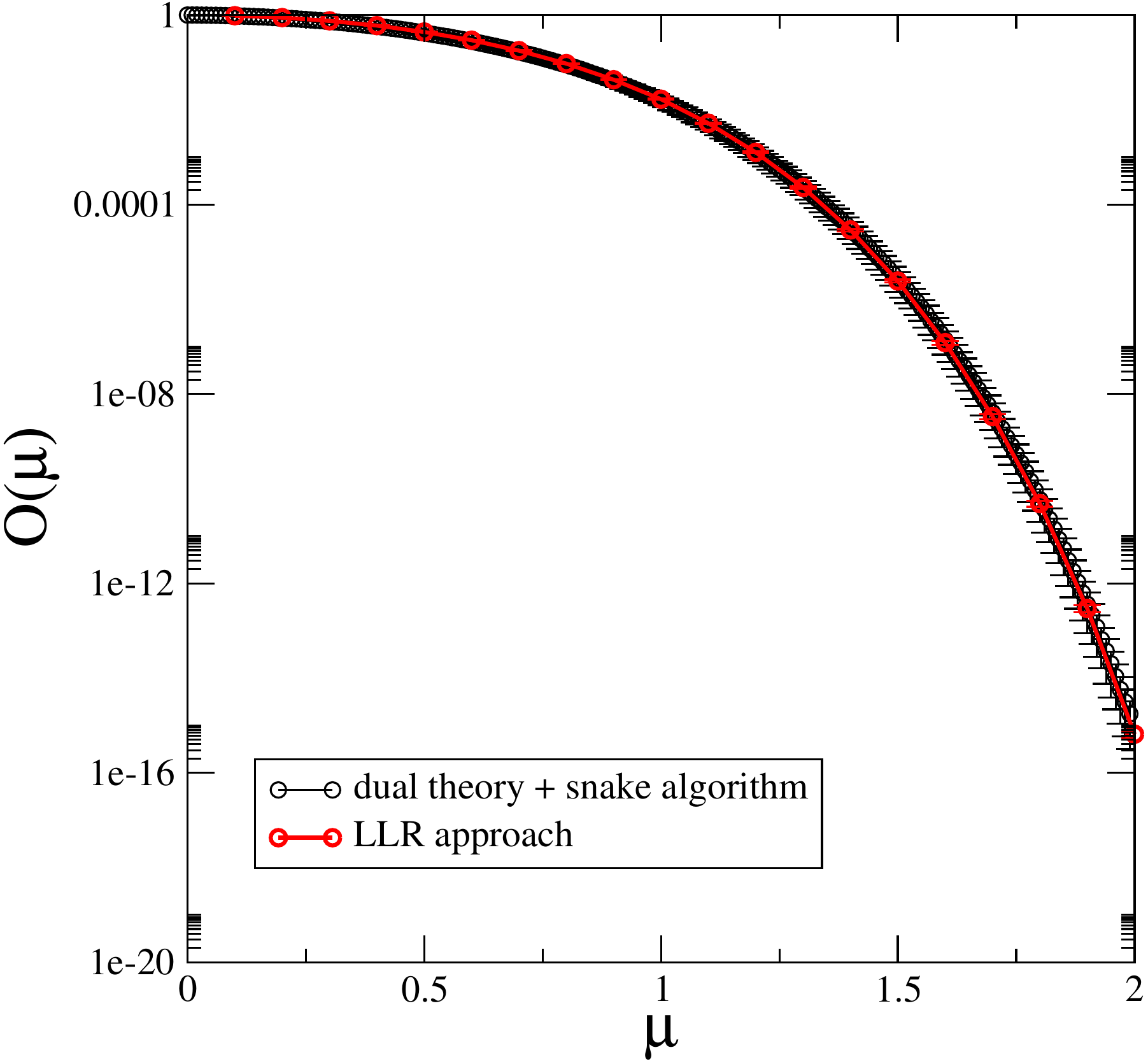} 
\caption{\label{fig:2} The phase factor calculated using the flux algorithm
  and the dual formulation (black symbols) and the LLR algorithm applied to
  the original theory with a strong sign problem (red symbols). $24^3$
  lattice, $\tau =0.1$ and $\kappa =0.01$. 
}
\end{figure}
Our aim will be to calculate $ \rho (n) $ with a precision of many 
of orders of magnitude such that a direct evaluation of (\ref{eq:18}) does
yield a statistically significant result despite cancellations. 
For this purpose, we follow~\cite{Langfeld:2012ah} and write 
\be 
\rho (n) \; = \; \prod_{i=0}^n \exp \{ - a_i \} \; . 
\label{eq:19} 
\en
We then define the ``double-bracket'' expectation values by: 
\bea 
\lb F \rb (a_n) &=& \frac{1}{\cal N} \sum _{\{\phi\}} \; F\left( 
\Delta N\left[\phi  \right]  \right) \; \theta  (\Delta N,n)\; \exp \{ a_n \}
\label{eq:20} \\ 
&& \exp \Bigl\{ S[\phi]  
\; + \; \kappa  \Bigl( 3N_0 [\phi] - V \Bigr) \, \hbox{cosh}(\mu)
 \Bigr\} \; , 
\nonumber \\ 
{\cal N} \; = \; \sum _{\{\phi\}} &&\theta (\Delta N,n) \; \exp \{ a_n \}
\label{eq:21} \\ 
&& \exp \Bigl\{ S[\phi]  
\; + \; \kappa  \Bigl( 3N_0 [\phi] - V \Bigr) \, \hbox{cosh}(\mu)
 \Bigr\} \; , 
\nonumber 
 \ena 
where $\theta (\Delta N,n) = 1 $ for $ \vert \Delta N[\phi] - n \vert \le 1 $
and $\theta (\Delta N,n)  = 0 $ otherwise. Note that these expectation values can
be calculated using standard Monte Carlo methods. The LLR key ingredient is
the observation that if the re-weighting factor $\exp \{ a \}$ is chosen 
correctly,  configurations with $\Delta N = n-1$, $\Delta N = n$ and 
$\Delta N = n+1$ possess the same probability. This yields a non-linear
equation to determine $a_n$:
\be 
\lb \Delta N \rb (a_n) \; = \; 0 \; . 
\label{eq:22}
\en 
It is the later equation which we solve using a Newton-Raphson iteration: 
\be 
a_n^{k+1} \; = \; a_n^k \; - \; \frac{ \lb \Delta N \rb (a_n^k) }{ 
\lb \Delta N^2 \rb (a_n^k)} \; . 
\label{eq:23}
\en
Details of the algorithm will be presented elsewhere. 
Once we have obtained
the coefficients $a_n$, we can reconstruct the density of states $\rho $ with
the help of (\ref{eq:19}). In practice, we have obtained $200$ independent
values for each of the $a_n$ with $n$ up to $5000$. Our result for $\rho_n$ is 
also shown in figure~\ref{fig:1}. Error bars are obtained using
the bootstrap method. We find an excellent agreement with the histogram method, but can
extend the observed range of $\rho $ to over $60$ orders of magnitude. 

%\medskip 
The phase factor $Q(\mu )$ can now obtained from (\ref{eq:6}) or, in the case
of the $Z_3$ spin model, from 
\be 
Q(\mu) \; = \; \frac{ 
\sum_{ n} \; \rho ( n) \; \cos \Bigl( \sqrt{3} \, \kappa 
\,   \hbox{sinh}(\mu) \; n \Bigr) }{ \sum_{ n} \; \rho ( n) } \; . 
\label{eq:30} 
\en
Error margins could once again be computed using bootstrap. However, we found it
advantageous to exploit the smoothness of $\ln \rho (n)$ and fit this function
to an even polynomial of degree $2p$: 
\be 
\ln \rho (n) \; = \; \sum _{k=0}^p c_k \, n^{2k} \; . 
\label{eq:31} 
\en 
In practice, we fitted polynomials of degree $2p = 2,4,6,8$ and found very
stable results with only the coefficients $c_0$ and $c_2$ significantly (within
bootstrap error bars) different from zero. 
After the extraction of the Taylor coefficients, the phase factor
(\ref{eq:30}) can be obtained ``semi-analytical'' to a high precision. 

Our numerical findings for $\rho$ are shown in figure~\ref{fig:1},
while our results for $Q(\mu)$ are summarised in
figure~\ref{fig:2}. Our density of states agrees with the density of
states extracted from the flux algorithm for all values for which the
latter method is effective (see figure~\ref{fig:1}, left panel, for an
example). The right 
panel of figure~\ref{fig:1} demonstrates the ability of our method to
determine the density of states over more than sixty orders of
magnitude. The correctness of this determination can be checked by
comparing our results for $Q(\mu)$ with results obtained with the flux
algorithm. Figure~\ref{fig:2} shows agreement over a wide range of
$\mu$, which determines a variation of $Q(\mu)$ over sixteen orders of
magnitude. A more detailed inspection shows that numerical results
(obtained using quadruple precision) found with the two methods are
always within errors. 

In conclusion, we have proposed an efficient {\em ab initio} approach 
that allows us to compute numerically observables affected by
strong cancellations in systems afflicted by a sign problem. The
methods consists of: a) a generalisation of the density of states; b)
a numerical determination of the generalised density of states using
the LLR algorithm; c) a polynomial interpolation of the density of
states; d) a semi-analytical determination of observables. This
strategy has been successfully probed for the $Z_3$ spin system, 
for which numerical results are available since its dual formualtion is real
and accessible by Monte Carlo methods. We have found that our method
reproduces results of the dual formulation over a wide range of chemical
potentials. The final goal of our programme would be tackling the sign
problem in QCD and in other real-world systems. In order to verify
the effectiveness of our method, studies of more complicated toy
models such as the O(2) system and the Bose gas at final temperature
are currently in progress. 

%\vskip 1mm
%\par\bigskip
\noindent {\bf Acknowledgments:} 
We thank Christof Gattringer, Jeff Greensite, Ydalia Mercado and Antonio
Rago for discussions. This work is supported by STFC under the DiRAC
framework. We are grateful for the support from the HPCC Plymouth,
where the numerical computations have been carried out. KL is supported by 
STFC (grant ST/L000350/1). BL is
supported by the Royal Society (grant UF09003) and by STFC (grant
ST/G000506/1). 

\bibliography{density_ym}{}

\begin{thebibliography}{18}%
\makeatletter
\providecommand \@ifxundefined [1]{%
 \@ifx{#1\undefined}
}%
\providecommand \@ifnum [1]{%
 \ifnum #1\expandafter \@firstoftwo
 \else \expandafter \@secondoftwo
 \fi
}%
\providecommand \@ifx [1]{%
 \ifx #1\expandafter \@firstoftwo
 \else \expandafter \@secondoftwo
 \fi
}%
\providecommand \natexlab [1]{#1}%
\providecommand \enquote  [1]{``#1''}%
\providecommand \bibnamefont  [1]{#1}%
\providecommand \bibfnamefont [1]{#1}%
\providecommand \citenamefont [1]{#1}%
\providecommand \href@noop [0]{\@secondoftwo}%
\providecommand \href [0]{\begingroup \@sanitize@url \@href}%
\providecommand \@href[1]{\@@startlink{#1}\@@href}%
\providecommand \@@href[1]{\endgroup#1\@@endlink}%
\providecommand \@sanitize@url [0]{\catcode `\\12\catcode `\$12\catcode
  `\&12\catcode `\#12\catcode `\^12\catcode `\_12\catcode `\%12\relax}%
\providecommand \@@startlink[1]{}%
\providecommand \@@endlink[0]{}%
\providecommand \url  [0]{\begingroup\@sanitize@url \@url }%
\providecommand \@url [1]{\endgroup\@href {#1}{\urlprefix }}%
\providecommand \urlprefix  [0]{URL }%
\providecommand \Eprint [0]{\href }%
\providecommand \doibase [0]{http://dx.doi.org/}%
\providecommand \selectlanguage [0]{\@gobble}%
\providecommand \bibinfo  [0]{\@secondoftwo}%
\providecommand \bibfield  [0]{\@secondoftwo}%
\providecommand \translation [1]{[#1]}%
\providecommand \BibitemOpen [0]{}%
\providecommand \bibitemStop [0]{}%
\providecommand \bibitemNoStop [0]{.\EOS\space}%
\providecommand \EOS [0]{\spacefactor3000\relax}%
\providecommand \BibitemShut  [1]{\csname bibitem#1\endcsname}%
\let\auto@bib@innerbib\@empty
%</preamble>
\bibitem [{\citenamefont {Creutz}(1980)}]{Creutz:1980zw}%
  \BibitemOpen
  \bibfield  {author} {\bibinfo {author} {\bibfnamefont {M.}~\bibnamefont
  {Creutz}},\ }\href {\doibase 10.1103/PhysRevD.21.2308} {\bibfield  {journal}
  {\bibinfo  {journal} {Phys.Rev.}\ }\textbf {\bibinfo {volume} {D21}},\
  \bibinfo {pages} {2308} (\bibinfo {year} {1980})}\BibitemShut {NoStop}%
%%CITATION = PHRVA,D21,2308;%%
\bibitem [{\citenamefont {Wilson}(1974)}]{Wilson:1974sk}%
  \BibitemOpen
  \bibfield  {author} {\bibinfo {author} {\bibfnamefont {K.~G.}\ \bibnamefont
  {Wilson}},\ }\href {\doibase 10.1103/PhysRevD.10.2445} {\bibfield  {journal}
  {\bibinfo  {journal} {Phys.Rev.}\ }\textbf {\bibinfo {volume} {D10}},\
  \bibinfo {pages} {2445} (\bibinfo {year} {1974})}\BibitemShut {NoStop}%
%%CITATION = PHRVA,D10,2445;%%
\bibitem [{\citenamefont {Gregory}\ \emph {et~al.}(2010)\citenamefont
  {Gregory}, \citenamefont {Davies}, \citenamefont {Follana}, \citenamefont
  {Gamiz}, \citenamefont {Kendall} \emph {et~al.}}]{Gregory:2009hq}%
  \BibitemOpen
  \bibfield  {author} {\bibinfo {author} {\bibfnamefont {E.}~\bibnamefont
  {Gregory}}, \bibinfo {author} {\bibfnamefont {C.}~\bibnamefont {Davies}},
  \bibinfo {author} {\bibfnamefont {E.}~\bibnamefont {Follana}}, \bibinfo
  {author} {\bibfnamefont {E.}~\bibnamefont {Gamiz}}, \bibinfo {author}
  {\bibfnamefont {I.}~\bibnamefont {Kendall}},  \emph {et~al.},\ }\href
  {\doibase 10.1103/PhysRevLett.104.022001} {\bibfield  {journal} {\bibinfo
  {journal} {Phys.Rev.Lett.}\ }\textbf {\bibinfo {volume} {104}},\ \bibinfo
  {pages} {022001} (\bibinfo {year} {2010})},\ \Eprint
  {http://arxiv.org/abs/0909.4462} {arXiv:0909.4462 [hep-lat]} \BibitemShut
  {NoStop}%
%%CITATION = ARXIV:0909.4462;%%
\bibitem [{\citenamefont {Foulkes}\ \emph {et~al.}(2001)\citenamefont
  {Foulkes}, \citenamefont {Mitas}, \citenamefont {Needs},\ and\ \citenamefont
  {Rajagopal}}]{RevModPhys.73.33}%
  \BibitemOpen
  \bibfield  {author} {\bibinfo {author} {\bibfnamefont {W.~M.~C.}\
  \bibnamefont {Foulkes}}, \bibinfo {author} {\bibfnamefont {L.}~\bibnamefont
  {Mitas}}, \bibinfo {author} {\bibfnamefont {R.~J.}\ \bibnamefont {Needs}}, \
  and\ \bibinfo {author} {\bibfnamefont {G.}~\bibnamefont {Rajagopal}},\ }\href
  {\doibase 10.1103/RevModPhys.73.33} {\bibfield  {journal} {\bibinfo
  {journal} {Rev. Mod. Phys.}\ }\textbf {\bibinfo {volume} {73}},\ \bibinfo
  {pages} {33} (\bibinfo {year} {2001})}\BibitemShut {NoStop}%
\bibitem [{\citenamefont {Aarts}\ and\ \citenamefont
  {James}(2012)}]{Aarts:2011zn}%
  \BibitemOpen
  \bibfield  {author} {\bibinfo {author} {\bibfnamefont {G.}~\bibnamefont
  {Aarts}}\ and\ \bibinfo {author} {\bibfnamefont {F.~A.}\ \bibnamefont
  {James}},\ }\href {\doibase 10.1007/JHEP01(2012)118} {\bibfield  {journal}
  {\bibinfo  {journal} {JHEP}\ }\textbf {\bibinfo {volume} {1201}},\ \bibinfo
  {pages} {118} (\bibinfo {year} {2012})},\ \Eprint
  {http://arxiv.org/abs/1112.4655} {arXiv:1112.4655 [hep-lat]} \BibitemShut
  {NoStop}%
%%CITATION = ARXIV:1112.4655;%%
\bibitem [{\citenamefont {Aarts}\ \emph {et~al.}(2013)\citenamefont {Aarts},
  \citenamefont {James}, \citenamefont {Pawlowski}, \citenamefont {Seiler},
  \citenamefont {Sexty} \emph {et~al.}}]{Aarts:2012ft}%
  \BibitemOpen
  \bibfield  {author} {\bibinfo {author} {\bibfnamefont {G.}~\bibnamefont
  {Aarts}}, \bibinfo {author} {\bibfnamefont {F.~A.}\ \bibnamefont {James}},
  \bibinfo {author} {\bibfnamefont {J.~M.}\ \bibnamefont {Pawlowski}}, \bibinfo
  {author} {\bibfnamefont {E.}~\bibnamefont {Seiler}}, \bibinfo {author}
  {\bibfnamefont {D.}~\bibnamefont {Sexty}},  \emph {et~al.},\ }\href {\doibase
  10.1007/JHEP03(2013)073} {\bibfield  {journal} {\bibinfo  {journal} {JHEP}\
  }\textbf {\bibinfo {volume} {1303}},\ \bibinfo {pages} {073} (\bibinfo {year}
  {2013})},\ \Eprint {http://arxiv.org/abs/1212.5231} {arXiv:1212.5231
  [hep-lat]} \BibitemShut {NoStop}%
%%CITATION = ARXIV:1212.5231;%%
\bibitem [{\citenamefont {Prokof'ev}\ and\ \citenamefont
  {Svistunov}(2001)}]{Prokof'ev:2001zz}%
  \BibitemOpen
  \bibfield  {author} {\bibinfo {author} {\bibfnamefont {N.}~\bibnamefont
  {Prokof'ev}}\ and\ \bibinfo {author} {\bibfnamefont {B.}~\bibnamefont
  {Svistunov}},\ }\href {\doibase 10.1103/PhysRevLett.87.160601} {\bibfield
  {journal} {\bibinfo  {journal} {Phys.Rev.Lett.}\ }\textbf {\bibinfo {volume}
  {87}},\ \bibinfo {pages} {160601} (\bibinfo {year} {2001})}\BibitemShut
  {NoStop}%
%%CITATION = PRLTA,87,160601;%%
\bibitem [{\citenamefont {Alet}\ \emph {et~al.}(2005)\citenamefont {Alet},
  \citenamefont {Lucini},\ and\ \citenamefont {Vettorazzo}}]{Alet:2004rh}%
  \BibitemOpen
  \bibfield  {author} {\bibinfo {author} {\bibfnamefont {F.}~\bibnamefont
  {Alet}}, \bibinfo {author} {\bibfnamefont {B.}~\bibnamefont {Lucini}}, \ and\
  \bibinfo {author} {\bibfnamefont {M.}~\bibnamefont {Vettorazzo}},\ }\href
  {\doibase 10.1016/j.cpc.2005.03.082} {\bibfield  {journal} {\bibinfo
  {journal} {Comput.Phys.Commun.}\ }\textbf {\bibinfo {volume} {169}},\
  \bibinfo {pages} {370} (\bibinfo {year} {2005})},\ \Eprint
  {http://arxiv.org/abs/hep-lat/0409156} {arXiv:hep-lat/0409156 [hep-lat]}
  \BibitemShut {NoStop}%
%%CITATION = HEP-LAT/0409156;%%
\bibitem [{\citenamefont {Mercado}\ and\ \citenamefont
  {Gattringer}(2012)}]{Mercado:2012ue}%
  \BibitemOpen
  \bibfield  {author} {\bibinfo {author} {\bibfnamefont {Y.~D.}\ \bibnamefont
  {Mercado}}\ and\ \bibinfo {author} {\bibfnamefont {C.}~\bibnamefont
  {Gattringer}},\ }\href {\doibase 10.1016/j.nuclphysb.2012.05.009} {\bibfield
  {journal} {\bibinfo  {journal} {Nucl.Phys.}\ }\textbf {\bibinfo {volume}
  {B862}},\ \bibinfo {pages} {737} (\bibinfo {year} {2012})},\ \Eprint
  {http://arxiv.org/abs/1204.6074} {arXiv:1204.6074 [hep-lat]} \BibitemShut
  {NoStop}%
%%CITATION = ARXIV:1204.6074;%%
\bibitem [{\citenamefont {Langfeld}(2013)}]{Langfeld:2013kno}%
  \BibitemOpen
  \bibfield  {author} {\bibinfo {author} {\bibfnamefont {K.}~\bibnamefont
  {Langfeld}},\ }\href {\doibase 10.1103/PhysRevD.87.114504} {\bibfield
  {journal} {\bibinfo  {journal} {Phys.Rev.}\ }\textbf {\bibinfo {volume}
  {D87}},\ \bibinfo {pages} {114504} (\bibinfo {year} {2013})},\ \Eprint
  {http://arxiv.org/abs/1302.1908} {arXiv:1302.1908 [hep-lat]} \BibitemShut
  {NoStop}%
%%CITATION = ARXIV:1302.1908;%%
\bibitem [{\citenamefont {Chandrasekharan}\ and\ \citenamefont
  {Li}(2011)}]{Chandrasekharan:2010iy}%
  \BibitemOpen
  \bibfield  {author} {\bibinfo {author} {\bibfnamefont {S.}~\bibnamefont
  {Chandrasekharan}}\ and\ \bibinfo {author} {\bibfnamefont {A.}~\bibnamefont
  {Li}},\ }\href {\doibase 10.1007/JHEP01(2011)018} {\bibfield  {journal}
  {\bibinfo  {journal} {JHEP}\ }\textbf {\bibinfo {volume} {1101}},\ \bibinfo
  {pages} {018} (\bibinfo {year} {2011})},\ \Eprint
  {http://arxiv.org/abs/1008.5146} {arXiv:1008.5146 [hep-lat]} \BibitemShut
  {NoStop}%
%%CITATION = ARXIV:1008.5146;%%
\bibitem [{\citenamefont {Wang}\ and\ \citenamefont
  {Landau}(2001)}]{Wang:2001ab}%
  \BibitemOpen
  \bibfield  {author} {\bibinfo {author} {\bibfnamefont {F.}~\bibnamefont
  {Wang}}\ and\ \bibinfo {author} {\bibfnamefont {D.~P.}\ \bibnamefont
  {Landau}},\ }\href {\doibase 10.1103/PhysRevLett.86.2050} {\bibfield
  {journal} {\bibinfo  {journal} {Phys. Rev. Lett.}\ }\textbf {\bibinfo
  {volume} {86}},\ \bibinfo {pages} {2050} (\bibinfo {year}
  {2001})}\BibitemShut {NoStop}%
\bibitem [{\citenamefont {Langfeld}\ \emph {et~al.}(2012)\citenamefont
  {Langfeld}, \citenamefont {Lucini},\ and\ \citenamefont
  {Rago}}]{Langfeld:2012ah}%
  \BibitemOpen
  \bibfield  {author} {\bibinfo {author} {\bibfnamefont {K.}~\bibnamefont
  {Langfeld}}, \bibinfo {author} {\bibfnamefont {B.}~\bibnamefont {Lucini}}, \
  and\ \bibinfo {author} {\bibfnamefont {A.}~\bibnamefont {Rago}},\ }\href
  {\doibase 10.1103/PhysRevLett.109.111601} {\bibfield  {journal} {\bibinfo
  {journal} {Phys.Rev.Lett.}\ }\textbf {\bibinfo {volume} {109}},\ \bibinfo
  {pages} {111601} (\bibinfo {year} {2012})},\ \Eprint
  {http://arxiv.org/abs/1204.3243} {arXiv:1204.3243 [hep-lat]} \BibitemShut
  {NoStop}%
%%CITATION = ARXIV:1204.3243;%%
\bibitem [{\citenamefont {Langfeld}\ and\ \citenamefont
  {Pawlowski}(2013)}]{Langfeld:2013xbf}%
  \BibitemOpen
  \bibfield  {author} {\bibinfo {author} {\bibfnamefont {K.}~\bibnamefont
  {Langfeld}}\ and\ \bibinfo {author} {\bibfnamefont {J.~M.}\ \bibnamefont
  {Pawlowski}},\ }\href {\doibase 10.1103/PhysRevD.88.071502} {\bibfield
  {journal} {\bibinfo  {journal} {Phys.Rev.}\ }\textbf {\bibinfo {volume}
  {D88}},\ \bibinfo {pages} {071502} (\bibinfo {year} {2013})},\ \Eprint
  {http://arxiv.org/abs/1307.0455} {arXiv:1307.0455 [hep-lat]} \BibitemShut
  {NoStop}%
%%CITATION = ARXIV:1307.0455;%%
\bibitem [{\citenamefont {Mercado}\ \emph {et~al.}(2011)\citenamefont
  {Mercado}, \citenamefont {Evertz},\ and\ \citenamefont
  {Gattringer}}]{Mercado:2011ua}%
  \BibitemOpen
  \bibfield  {author} {\bibinfo {author} {\bibfnamefont {Y.~D.}\ \bibnamefont
  {Mercado}}, \bibinfo {author} {\bibfnamefont {H.~G.}\ \bibnamefont {Evertz}},
  \ and\ \bibinfo {author} {\bibfnamefont {C.}~\bibnamefont {Gattringer}},\
  }\href {\doibase 10.1103/PhysRevLett.106.222001} {\bibfield  {journal}
  {\bibinfo  {journal} {Phys.Rev.Lett.}\ }\textbf {\bibinfo {volume} {106}},\
  \bibinfo {pages} {222001} (\bibinfo {year} {2011})},\ \Eprint
  {http://arxiv.org/abs/1102.3096} {arXiv:1102.3096 [hep-lat]} \BibitemShut
  {NoStop}%
%%CITATION = ARXIV:1102.3096;%%
\bibitem [{\citenamefont {Karsch}\ and\ \citenamefont
  {Wyld}(1985)}]{Karsch:1985cb}%
  \BibitemOpen
  \bibfield  {author} {\bibinfo {author} {\bibfnamefont {F.}~\bibnamefont
  {Karsch}}\ and\ \bibinfo {author} {\bibfnamefont {H.}~\bibnamefont {Wyld}},\
  }\href {\doibase 10.1103/PhysRevLett.55.2242} {\bibfield  {journal} {\bibinfo
   {journal} {Phys.Rev.Lett.}\ }\textbf {\bibinfo {volume} {55}},\ \bibinfo
  {pages} {2242} (\bibinfo {year} {1985})}\BibitemShut {NoStop}%
%%CITATION = PRLTA,55,2242;%%
\bibitem [{\citenamefont {DeGrand}\ and\ \citenamefont
  {DeTar}(1983)}]{DeGrand:1983fk}%
  \BibitemOpen
  \bibfield  {author} {\bibinfo {author} {\bibfnamefont {T.~A.}\ \bibnamefont
  {DeGrand}}\ and\ \bibinfo {author} {\bibfnamefont {C.~E.}\ \bibnamefont
  {DeTar}},\ }\href {\doibase 10.1016/0550-3213(83)90536-9} {\bibfield
  {journal} {\bibinfo  {journal} {Nucl.Phys.}\ }\textbf {\bibinfo {volume}
  {B225}},\ \bibinfo {pages} {590} (\bibinfo {year} {1983})}\BibitemShut
  {NoStop}%
%%CITATION = NUPHA,B225,590;%%
\bibitem [{\citenamefont {de~Forcrand}\ \emph {et~al.}(2001)\citenamefont
  {de~Forcrand}, \citenamefont {D'Elia},\ and\ \citenamefont
  {Pepe}}]{deForcrand:2000fi}%
  \BibitemOpen
  \bibfield  {author} {\bibinfo {author} {\bibfnamefont {P.}~\bibnamefont
  {de~Forcrand}}, \bibinfo {author} {\bibfnamefont {M.}~\bibnamefont {D'Elia}},
  \ and\ \bibinfo {author} {\bibfnamefont {M.}~\bibnamefont {Pepe}},\ }\href
  {\doibase 10.1103/PhysRevLett.86.1438} {\bibfield  {journal} {\bibinfo
  {journal} {Phys.Rev.Lett.}\ }\textbf {\bibinfo {volume} {86}},\ \bibinfo
  {pages} {1438} (\bibinfo {year} {2001})},\ \Eprint
  {http://arxiv.org/abs/hep-lat/0007034} {arXiv:hep-lat/0007034 [hep-lat]}
  \BibitemShut {NoStop}%
\end{thebibliography}%

\end{document}